\documentclass[%
aip,
amsmath,amssymb,
reprint,
]{revtex4-2}
\usepackage{braket}
\usepackage{graphicx}
\usepackage{booktabs}
\usepackage{siunitx}
\usepackage{bm}
\usepackage{xcolor}
\usepackage[normalem]{ulem}
\usepackage[utf8]{inputenc}
\usepackage[T1]{fontenc}
\usepackage{mathptmx}

\begin{document}

\title{A Highly Nondegenerate Photon-Pair Source for Detector-Matched Quantum Key Distribution}

\author{Sander van Haagen}
\affiliation{Qubitrium B.V., Delft, Netherlands}
\author{Bora Ulu}
\affiliation{Qubitrium A.S., Istanbul, T\"urkiye}
\author{Alper Ozulker}
\affiliation{Qubitrium A.S., Istanbul, T\"urkiye}
\author{Wardah Mahmood}
\affiliation{Qubitrium A.S., Istanbul, T\"urkiye}
\author{Michal Krelina}
\affiliation{Qubitrium B.V., Delft, Netherlands}
\author{Utku Tefek}
\affiliation{Qubitrium B.V., Delft, Netherlands}
\author{Kadir Durak}
\affiliation{Qubitrium A.S., Istanbul, T\"urkiye}
\affiliation{Electrical and Electronics Engineering Department, Özyeğin University, Istanbul, Turkey.}

\begin{abstract}
For a commercial entanglement-distribution system over telecom fibre, the single-photon detectors are the largest part of the hardware cost. In entanglement-based quantum key distribution (QKD), both photons of each pair must be detected.
With two telecom photons, each arm needs either an InGaAs single-photon avalanche diode (SPAD) of low efficiency or a superconducting nanowire single-photon detector (SNSPD) with its own cryostat. We study a highly nondegenerate photon-pair source that needs a telecom detector on one arm only. Two periodically poled KTiOPO$_4$ (ppKTP) crystals pumped at 405~nm emit pairs at 548 and 1550~nm. The 548~nm photon is detected at the source node by a Si-SPAD with 82\% efficiency, and only the 1550~nm photon travels through fibre. A first, unoptimised module in a polarisation-correlated configuration generates about $1.5\times10^{6}$ pairs/s per mW of pump power. It collects 13\% of the pairs on the 548~nm arm and 23\% on the 1550~nm arm. We use these values in an asymptotic BBM92 model with four detectors per receiver. The generic $2\times1550$~nm links use a bright published source at the same pump power. At 50~km, a link with this source and one SNSPD then costs about 25\% more per secret bit than a generic link with one SNSPD, and about six times as much as a generic link with two SNSPDs. Under optimistic target values, with 52\% collection and twice the brightness, the link with one SNSPD costs about 40\% as much per bit as the two-SNSPD link. A link without any cryostat then costs about 70\% less per bit than the generic link with one SNSPD. If every link runs at its own best pump power, the two-SNSPD link is the cheapest again, at about two thirds of the cost of the target link with one SNSPD. 
Beyond QKD, the source can also serve pair-based clock synchronisation and the calibration of telecom detectors. These uses need only the time correlation of the pairs, not their entanglement. 
The entangled configuration of the source is still under development, and all key rates assume this configuration.
\end{abstract}

\maketitle

\section{Introduction}\label{sec:intro}
This work aims at a commercially viable system that distributes entangled photon pairs over telecom fibre. Its first application is entanglement-based quantum key distribution (QKD). A commercial QKD link must deliver key at a low cost per secret bit. In the generic links that we model, the telecom detectors account for 77 to 92\% of the hardware cost (Table~\ref{tab:cost}).

QKD lets two parties share a secret key whose security rests on quantum physics~\cite{pirandola2020advances,xu2020secure}. In the entanglement-based protocol BBM92~\cite{bennett1992quantum}, a simplified form of the Ekert protocol~\cite{ekert1991quantum}, a source emits photon pairs and each user measures one photon of every pair. A key bit comes only from a coincidence, in which both photons are detected. The key rate therefore scales with the product of the two detection efficiencies.

At the telecom wavelength of 1550~nm, a high detection efficiency comes at a high price. Free-running InGaAs/InP single-photon avalanche diodes (SPADs) operate without cryogenics, but they detect only a fraction of the photons and stay blind for microseconds after each detection. The commercial detector used in this work runs at 20\% efficiency, and a laboratory device reaches 40\% at a 10~$\mu$s dead time~\cite{xu2024freerunning}. Superconducting nanowire single-photon detectors (SNSPDs) reach 98\% in the laboratory~\cite{reddy2020superconducting}, but they operate at cryogenic temperatures and need a cryostat~\cite{natarajan2012superconducting}. With two telecom photons, every arm with an SNSPD needs its own cryostat, and every arm with an InGaAs SPAD misses most of its photons.

A highly nondegenerate photon-pair source needs a telecom detector on one arm only. One photon of each pair lies in the visible or near-infrared range, where Si-SPADs are efficient without cryogenics. The other photon lies in the telecom band, where fibre loss is low. The visible photon is detected at the source node, and only the telecom photon travels through fibre~\cite{hentschel2009three,telecom_eps_2025}. A link with SNSPDs then needs one cryostat instead of two, and a link without cryostats still has an efficient detector on one arm.
We call such a design detector-matched, because the wavelength of each photon suits its detector or its channel.

Sources of this kind have been built at 810 and 1550~nm, with two crossed crystals~\cite{pelton2004bright} or in a Sagnac interferometer~\cite{hentschel2009three,peng2025nondegenerate}, and recently at 680 and 1550~nm~\cite{telecom_eps_2025}. A commercial BBM92 system already uses a source at 810 and 1550~nm, with Si-SPADs at the source node and InGaAs SPADs at the remote node~\cite{qoj_elvis}. Our source emits at 548 and 1550~nm from a 405~nm pump. At 548~nm, the Si-SPAD of this work detects 82\% of the photons, close to its peak efficiency (Sec.~\ref{sec:design}). The first module was built in a polarisation-correlated configuration, with two crossed crystals and no phase compensation. We expect this configuration to produce no entanglement and therefore no secret key. All key rates in this paper therefore assume the entangled configuration of the source, which is under development.

Highly nondegenerate pairs are also useful beyond QKD. One photon can be detected by a silicon detector or camera, while its partner probes a sample or travels through fibre. Imaging and spectroscopy with undetected photons use this split to measure at a wavelength that is hard to detect~\cite{lemos2014quantum,kalashnikov2016infrared}. Telecom-heralded quantum memories store the photon that matches a memory transition and send its telecom partner through fibre~\cite{lagorivera2021telecom}. With a source at 810 and 1550~nm, a quantum state has been teleported from 810 to 1550~nm~\cite{peng2025nondegenerate}. Clock synchronisation with photon pairs uses the tight time correlation of the two photons~\cite{krelina2026quantum}. It has been shown with pairs at 810~nm~\cite{lee2019symmetrical,spiess2023clock}. Here we focus on QKD, and Section~\ref{sec:conclusions} returns to the other uses.

In this paper, we examine what the design achieves with this first module and what it needs to compete. Section~\ref{sec:source} describes the source and characterises the module. Section~\ref{sec:design} uses the measured values in a BBM92 model with detector dead time and afterpulsing. It compares two links built on this source with three links built on a generic $2\times1550$~nm source, in secret-key rate and in cost per secret bit. The comparison covers the module as measured and a target module. Section~\ref{sec:security} states the conditions for security, and Section~\ref{sec:conclusions} concludes.

\section{Source description and parameters}\label{sec:source}

The source is a type-0 ($e\!\to\!ee$) spontaneous parametric down-conversion source built on periodically poled KTiOPO$_4$ (ppKTP), pumped by a single-frequency laser at \SI{405}{nm} with a linewidth below \SI{160}{MHz} and emitting collinear pairs at \SI{548}{nm} and \SI{1550}{nm}. We prefer ppKTP over ppLN for a UV pump, because it has a higher damage threshold and is more robust to temperature drift. Type-0 quasi-phase matching is brighter than type-II but it has tighter thermal and spectral tolerances. 
In each crystal, the pump, signal and idler are polarised along the crystal $z$ axis. The two crystals have orthogonal $z$ axes, vertical and horizontal in our laboratory frame, so a diagonally polarised pump drives both. The emission wavelengths are set by the crystal temperatures, each stabilised to $\pm0.1^\circ$C.
Both source configurations below use the same crystal type and phase matching.

\begin{figure*}[htb!]
    \centering
    \includegraphics[width=\textwidth,trim=60 0 25 65,clip]{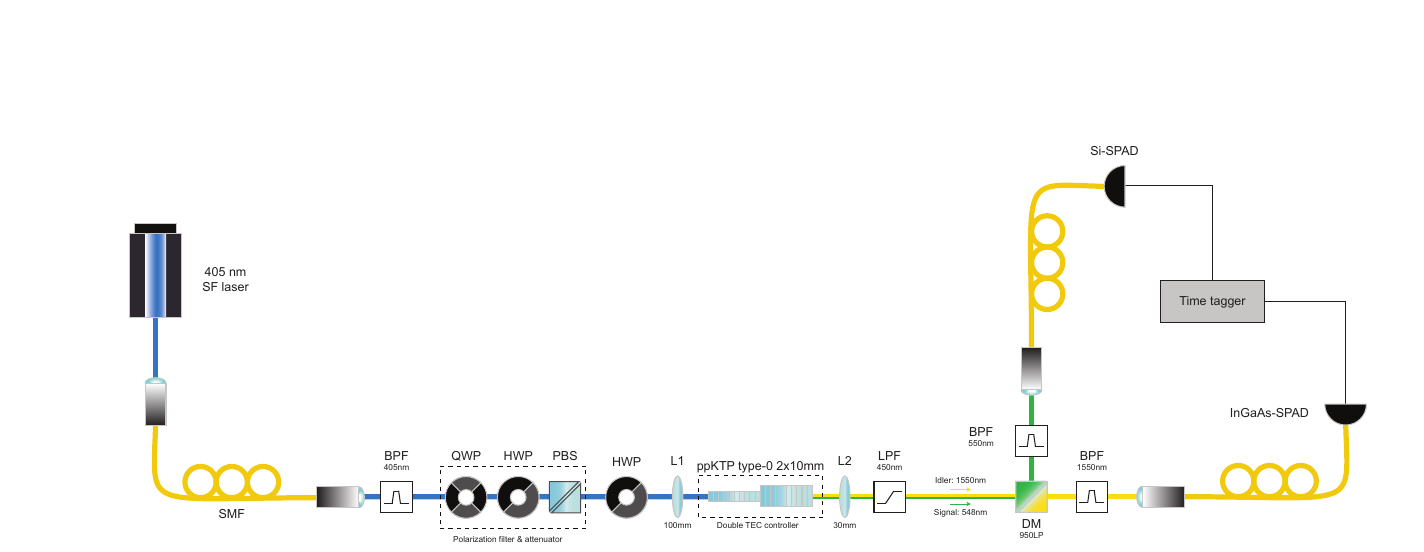}
\caption{\textbf{Correlated configuration of the source, as measured in Table~\ref{tab:source}.} A 405~nm single-frequency (SF) laser is coupled into a single-mode fibre (SMF). A 405~nm bandpass filter (BPF), a polarisation filter and attenuator (QWP, HWP, PBS) and a second HWP set the power and the diagonal polarisation of the pump. Lens $L_1$ ($f=100$~mm) focuses the pump into two 10~mm type-0 ppKTP crystals with orthogonal axes, each on its own thermoelectric temperature controller (TEC). Lens $L_2$ ($f=30$~mm) collimates the pairs, and a long-pass filter (LPF, 450~nm) removes the pump. A dichroic mirror (DM, 950~nm long-pass) sends the 548~nm signal and the 1550~nm idler through bandpass filters into single-mode fibres. A Si-SPAD and an InGaAs SPAD detect the photons, and a time tagger records the coincidences. Each pair is produced as HH or as VV, with no stable phase between the two, so this configuration is expected to give polarisation correlation without entanglement.} 
    \label{fig:nondeg_source_setup}
\end{figure*}

We designed the source in two configurations. In the \emph{correlated} configuration (Fig.~\ref{fig:nondeg_source_setup}), 
two ppKTP crystals are designed to produce photon pairs that are correlated in polarisation but not entangled.
The two crystals have orthogonal axes and sit one after the other. A diagonally polarised pump passes through both once. In the \emph{entangled} configuration, two such crystals sit inside a Sagnac interferometer (Fig.~\ref{fig:scenarios}), and the pairs carry polarisation entanglement. All measurements in this paper come from the correlated configuration. The entangled configuration is under development, with a target fidelity above 95\%. In the correlated configuration, we balanced the H and V populations during alignment with a polarisation analyser (half-wave plate, quarter-wave plate and polarising beamsplitter). The measured single-arm populations are
\begin{equation}
\rho = p_H\ket{H}\bra{H} + p_V\ket{V}\bra{V}, \quad p_H,p_V = 0.50 \pm 0.025.
\label{eq:rho_source}
\end{equation}
Following the design of this configuration, we treat the two-photon state as the separable mixture $\rho_{AB} = p_H\ket{HH}\bra{HH} + p_V\ket{VV}\bra{VV}$. A maximally entangled state needs equal populations and, in addition, mutually coherent amplitudes. Without this coherence, the state is not entangled. Coincidence data in the H/V basis alone cannot show this difference. The mixture passes a correlation test in the rectilinear basis but fails it in the diagonal basis (Sec.~\ref{sec:security}). Therefore it cannot violate a Bell inequality.

The measurements below come from a first, small-scale module. Its development focused on the crystals and on the miniaturised layout. The collection optics, and with them the heralding efficiency, are not yet optimised. The pump beam is spatially multimode and not filtered, and the mode matching between the pump and the collection modes is not optimised. The small housing, 25$\times$12.3$\times$12~cm, also limits the alignment. On the optical bench, a pump in a single spatial mode doubled the brightness. 
We expect it to raise the heralding efficiency as well. We could not measure this gain, because the bench measurement used different detectors.
The efficiencies below therefore describe this module and not the limit of the design. The link results that use them in Sec.~\ref{sec:design} are preliminary. Section~\ref{sec:collection} shows how those results change as the collection improves.

\begin{figure}[htb!]
    \centering
    \includegraphics[width=0.99\linewidth]{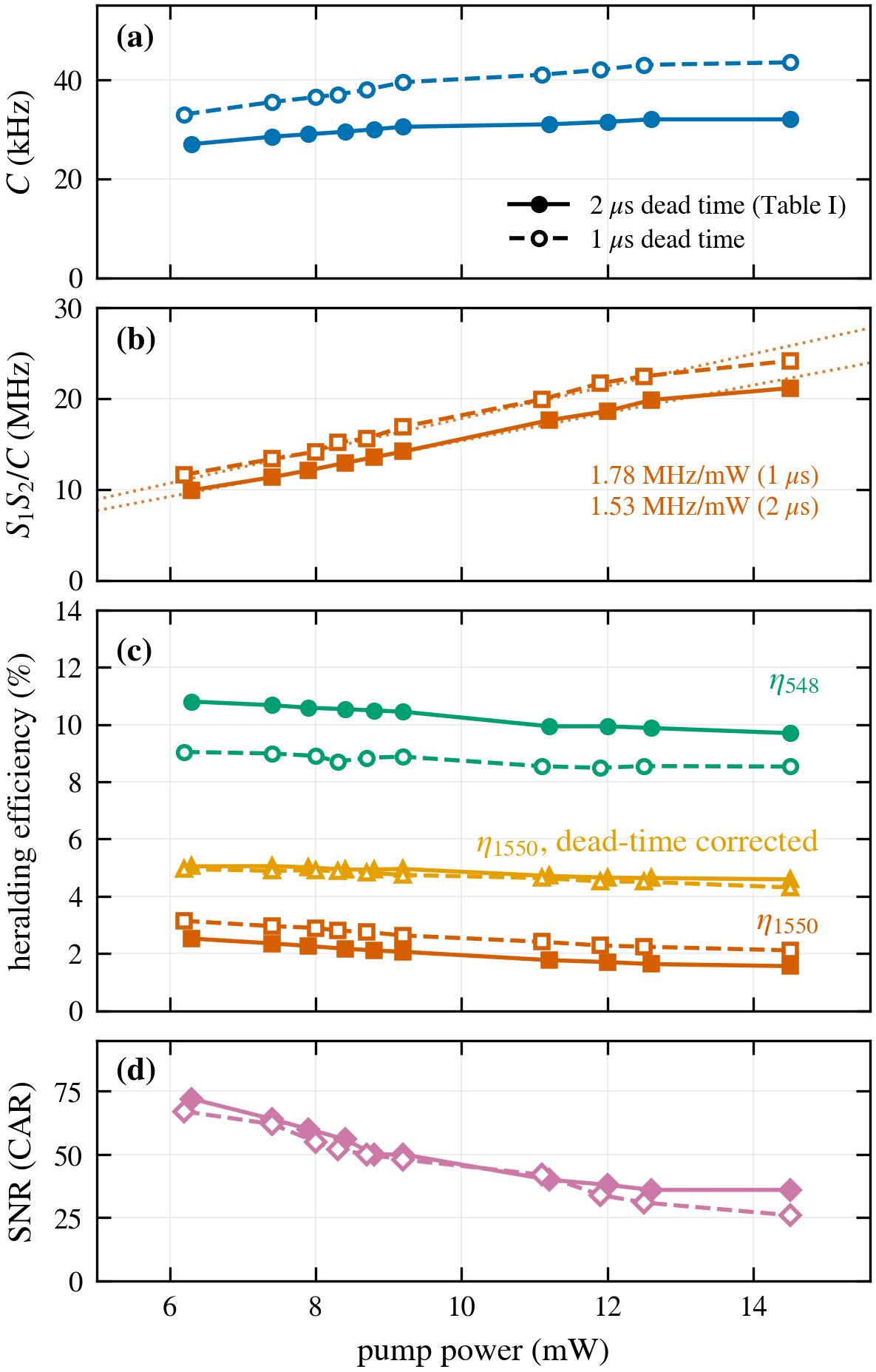}
\caption{\label{fig:source_scaling}\textbf{Source scaling with pump power.} Two runs on the same source with the same InGaAs detector: 2~$\mu$s dead time (filled markers, solid lines; Table~\ref{tab:source}) and 1~$\mu$s dead time (open markers, dashed lines). (a) The coincidence rate saturates, more strongly at the longer dead time. (b) The pair-generation rate (PGR) $S_1S_2/C$ stays nearly linear, at 1.53~MHz/mW (2~$\mu$s) and 1.78~MHz/mW (1~$\mu$s); dotted lines are fits through the origin. The higher slope at 1~$\mu$s is consistent with afterpulsing, which adds unpaired counts to $S_2$. (c) Both single-arm heralding efficiencies fall with pump power; triangles show $\eta_{1550}$ corrected for the dead time, which agrees between the two runs. (d) Coincidence-peak signal-to-noise ratio, taken as the coincidence-to-accidental ratio. It bounds the accessible multi-pair contamination. The implied coincidence window is 1.2--1.6~ns across the two runs.}
\end{figure}

Table~\ref{tab:source} lists the measured singles, coincidences and derived figures of merit of the correlated configuration at three pump powers. Figure~\ref{fig:source_scaling} shows the full scan, together with a second run at a shorter dead time. The 548~nm arm used a fibre-coupled Si-SPAD (made by Qubitrium) and the 1550~nm arm used a free-running InGaAs-SPAD (ID~Quantique ID~Qube NIR FR) operated at 20\% photon-detection efficiency (PDE) with a 2~$\mu$s dead time. Their dark-count rates are about 5~kHz (Si) and 1.5~kHz (InGaAs, at $-40^\circ$C). Dark counts are therefore at most about 0.6\% of the singles on either arm, and we neglect them. 

Three features matter for Sec.~\ref{sec:design}. First, the heralding efficiencies are low and unequal: 10.8\% at 548~nm against 2.5\% at 1550~nm at 6.3~mW. Both fall with pump power, mainly because the detectors saturate. Second, the coincidence rate is sublinear. It rises by a factor of 1.2 for a 2.3-fold pump increase, mainly because the free-running InGaAs detector is deep in its dead time. Its live fraction falls from 0.50 to 0.34 across the scan. 

We correct each heralding efficiency for the dead time of its own detector, by dividing it by that detector's live fraction. At 14.5~mW, this raises the 1550~nm efficiency from 1.6\% to 4.6\% and the 548~nm efficiency from 9.7\% to 11.1\%. After the correction, the two efficiencies change little across the scan: from 5.0\% to 4.6\% at 1550~nm and from 11.5\% to 11.1\% at 548~nm. 
In relative terms, the fall shrinks from 38\% to 9\% at 1550~nm and from 10\% to 4\% at 548~nm. Detector saturation therefore explains most of the relative fall at 1550~nm, but only about 60\% of it at 548~nm.The remainder at 548~nm would disappear for a Si-SPAD dead time of about 90~ns instead of the 60~ns of the detector datasheet~\cite{qubitrium_qdcv}.

Third, the pair-generation rate (PGR), estimated as $S_1S_2/C$, needs no such correction. The live fractions are the same in the two singles rates and in the coincidence rate, so they cancel in the ratio. The PGR is therefore nearly linear in the pump power, at 1.5~MHz/mW.

The two dead-time settings test the detector model of Sec.~\ref{sec:design} on one unchanged source. The 548~nm singles of the two runs agree to within 3\%. After the dead-time correction, the 1550~nm arm efficiency agrees between the runs to about 6\%: 4.3\% at 1~$\mu$s against 4.6\% at 2~$\mu$s, at 14.5~mW. The 548~nm efficiency $C/S_2$ should not depend on the InGaAs dead time at all, yet it is 12--17\% lower at 1~$\mu$s. At the shorter dead time, therefore, some of the 1550~nm counts have no partner photon. Afterpulsing is the likely cause, because a longer dead time suppresses it. The same excess raises the slope of the PGR from 1.53 to 1.78~MHz/mW. Table~\ref{tab:source} and the model of Sec.~\ref{sec:design} therefore use the 2~$\mu$s run. Some afterpulsing probably remains at 2~$\mu$s, so its $\eta_{548}$ is still a lower bound, and its pair rate $S_1S_2/C$ is an upper bound by the same factor. In the current parameter set of Sec.~\ref{sec:design}, the two errors cancel in the rate of true coincidences, which depends on their product. We do not correct for it, because the afterpulse probability of the detector at these settings was not measured. The values in Table~\ref{tab:source} are the registered rates, with no correction.

Ref.~\cite{telecom_eps_2025} reports a source of the same type, with its visible arm at 680~nm. Its heralding efficiencies are higher than ours: about 34\% on the visible arm against our 10.8\%, and 18\% on the telecom arm against our 2.5\%. Ref.~\cite{telecom_eps_2025} labels these values as the idler (34\%) and the signal (18\%) heralding efficiencies. The 34\% value stays the same when the telecom SNSPD is replaced by an InGaAs SPAD with 20\% PDE, so it describes the detection of the visible photon.

On the telecom arm, most of this gap comes from the detector. Ref.~\cite{telecom_eps_2025} detects the telecom photon with an SNSPD at 35\% PDE, while we use an InGaAs SPAD at 20\% PDE that is deep into its dead time. With an InGaAs SPAD (20\% PDE, 20~$\mu$s dead time), the same reference measures a telecom-arm efficiency that falls from about 2.1\% to below 0.5\% as the pump power rises, which is of the same order as ours. On the visible arm, the gap comes from the collection and from afterpulsing. Afterpulsing adds unpaired counts to $S_2$ and so lowers $\eta_{548}$. Once the detector efficiencies are divided out, our module collects 0.135 of the pairs on the 548~nm arm and 0.230 on the 1550~nm arm (Table~\ref{tab:params}). For Ref.~\cite{telecom_eps_2025} we estimate about 0.52 on both arms, from its heralding efficiencies, the 35\% PDE of its SNSPD and an assumed 65\% PDE of its Si-SPAD at 680~nm. This value is close to the 66\% measured for an SAP500 at 676~nm~\cite{stipcevic2013sap500}.

The two sources also use different compromises between brightness and heralding. A loosely focused pump improves the coupling into a single-mode fibre but lowers the brightness~\cite{telecom_eps_2025}. Ref.~\cite{telecom_eps_2025} chose a loose focus for high heralding and reports a PGR of 0.56~MHz/mW. Our module is brighter: its PGR is 1.53~MHz/mW, about 2.7 times higher.

\begin{table}[htb!]
\caption{\label{tab:source}Measured performance of the correlated configuration at three pump powers, with the 1550~nm detector at a 2~$\mu$s dead time. $S_1$, $S_2$ and $C$ are the raw registered rates on the 548~nm arm, the 1550~nm arm and in coincidence. $\eta_{548}=C/S_2$ and $\eta_{1550}=C/S_1$ are the heralding efficiencies of the two arms, $\eta_{\rm sym}=C/\sqrt{S_1S_2}$, and the pair-generation rate (PGR) is estimated as $S_1S_2/C$. SNR is the coincidence-peak signal-to-noise ratio, equivalent to the coincidence-to-accidental ratio. }
\begin{ruledtabular}
\begin{tabular}{l c c c}
Pump (mW) & 6.3 & 9.2 & 14.5 \\
\colrule
$S_1$ (548~nm, kHz)      & 1070  & 1480 & 2050 \\
$S_2$ (1550~nm, kHz)     & 250  & 292  & 330  \\
$C$ (kHz)                & 27.0 & 30.5 & 32.0 \\
$\eta_{548}$             & 10.8\% & 10.4\% & 9.7\% \\
$\eta_{1550}$            & 2.5\% & 2.1\% & 1.6\% \\
$\eta_{\rm sym}$         & 5.2\% & 4.6\% & 3.9\% \\
$S_1S_2/C$ (MHz)         & 9.9 & 14.2 & 21.1 \\
SNR (CAR)                & 72 & 50 & 36 \\
\end{tabular}
\end{ruledtabular}
\end{table}

\begin{figure*}[t]
    \centering
    \includegraphics[width=\textwidth]{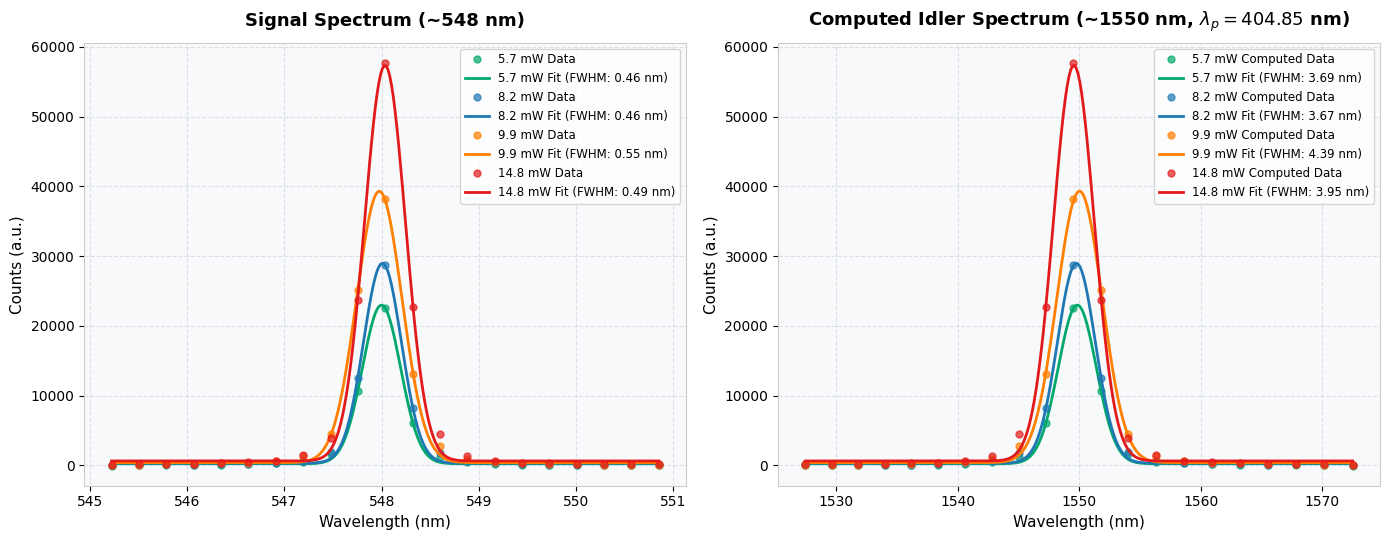}
    \caption{\label{fig:source_spectrum}\textbf{Signal and idler spectra.} Left: measured 548~nm spectrum at four pump powers, 5.7--14.8~mW. Right: 1550~nm idler spectrum derived from it by energy conservation with the \SI{405}{nm} pump.}
\end{figure*}

The 548~nm signal spectrum (Fig.~\ref{fig:source_spectrum}) was measured on a grating spectrometer. The 1550~nm idler spectrum follows from it by energy conservation with the single-frequency pump at 404.85~nm. 
The fitted widths are 0.46--0.55~nm at 548~nm and 3.7--4.4~nm at 1550~nm. In frequency both are 460--550~GHz. The two widths agree by construction, because we computed the idler spectrum and did not measure it. The agreement is therefore not an independent check. Both are limited by the spectrometer resolution. The spectrometer samples the signal every 0.28~nm, so the measured width spans fewer than two pixels. The phase-matching calculation predicts 0.2 and 1.5~nm, or about \SI{200}{GHz}.

Both configurations use a diagonally polarised pump, so each pair is produced either as HH or as VV. In the correlated configuration, the relative phase between the HH and VV amplitudes is not compensated. This phase varies across the broad emission spectrum, and the two amplitudes are also offset in time. The off-diagonal terms of the density matrix are therefore expected to average out, so that the pairs form the separable mixture $\rho_{AB}$ given above. We have not measured this. Entanglement needs a relative phase that is the same across the whole spectrum. In the entangled configuration, the two crystals sit inside a Sagnac interferometer (Fig.~\ref{fig:scenarios}). A polarising beamsplitter splits the pump into the two directions of a common-path loop. The two down-conversion amplitudes therefore occupy the same spatial mode, are intrinsically indistinguishable and carry a passively stable relative phase. The output is the Bell state $(\ket{HH}+e^{i\varphi}\ket{VV})/\sqrt2$.

The loop optics reduce the optical transmission by about 15\% relative to the correlated configuration. This loss enters the arm efficiencies of Sec.~\ref{sec:design}. This is a conservative value. With carefully chosen components, we expect a loss of about 10\%, which remains to be measured. The model applies this loss to each photon, so a coincidence carries $\eta_{\rm loop}^2$ [Eq.~\eqref{eq:armeff}].

\section{Detector-matched link designs}\label{sec:design}

\subsection{The physical asymmetry being exploited}
Our analysis rests on two facts about detectors. At \SI{1550}{nm} there is a sharp trade-off between efficiency and cost. A commercial free-running InGaAs/InP SPAD typically runs at 20--25\% PDE, while a SNSPD reaches $\geq$80\% (up to a measured \num{98.0(5)}\%~\cite{reddy2020superconducting}). However, the SNSPD needs a closed-cycle cryostat at about 0.8--2.5~K. At about 548~nm, by contrast, a silicon SPAD is already efficient: the Qubitrium SAP500 silicon SPAD reaches a measured $82\%$ PDE, comparable to a 1550~nm SNSPD but with Peltier cooling instead of a cryostat. Its detection-efficiency curve peaks at $83.5\%$ near 530~nm and stays between about 81\% and 83.5\% across 500--560~nm, so the exact wavelength of the visible arm within this band does not change the PDE.

In a BBM92 entanglement link~\cite{bennett1992quantum,waks2002security} both photons of a pair must be detected in coincidence, so the rate is set by the product of the two arm efficiencies. A symmetric $2\times$\SI{1550}{nm} link therefore needs two telecom-grade detectors, and the coincidence rate carries the 1550~nm detection efficiency twice (the $\eta^2$ detector product). Each of its two arms needs either a cryostat or an InGaAs SPAD of low efficiency. The highly nondegenerate source breaks this symmetry: it replaces one telecom detector with a room-temperature (with Peltier cooling) Si detector at 548~nm of about 80\% efficiency, so only the remaining 1550~nm arm carries the telecom-detector penalty. The asymmetry concerns detector efficiency and cryostat count. It does not concern channel loss. Both designs place the source at one node, so one arm is local and only one arm travels through fibre. In the loss-limited regime, both designs therefore scale with distance like the channel transmittance. A saturated detector on the travelling arm flattens this dependence at short distance (Sec.~\ref{sec:results}).

\subsection{Link scenarios}
We compare five link configurations (Table~\ref{tab:scenarios}), labelled by source family: \textbf{D} for the highly nondegenerate source measured here and \textbf{G} for a generic comparator source. We use this source for BBM92 in two detector variants: with an SNSPD (D1) or a free-running InGaAs SPAD (D2) on the travelling 1550~nm arm. The generic comparators run the same protocol with a symmetric $2\times$\SI{1550}{nm} entangled-photon source (EPS) and need a telecom detector on both arms. Their brightness and heralding efficiency are those of a bright published source in one 100~GHz channel~\cite{neumann2022experimental} (Sec.~\ref{sec:params}). We combine this source with three detector budgets: two SNSPDs (G1), one SNSPD and one InGaAs SPAD (G3), and two InGaAs SPADs (G2). These budgets need two, one and zero cryostats, respectively. G3 matters most for the comparison. It is the only link in the grid with one cryostat and without a visible arm, so it is the direct competitor to D1. Its InGaAs detectors sit on the travelling arm, where the channel loss lowers their count rate. 

We compare only pair sources. Weak-coherent-pulse links, such as decoy-state BB84, are a different class of transmitter. A gain-switched laser runs at a clock rate of 1~GHz and leads on raw throughput, whereas an SPDC source is limited by its pump and is scored per emitted pair. We leave those links out of scope. Section~\ref{sec:security} compares them qualitatively.

\begin{figure*}[t]
    \centering
    \includegraphics[width=\textwidth]{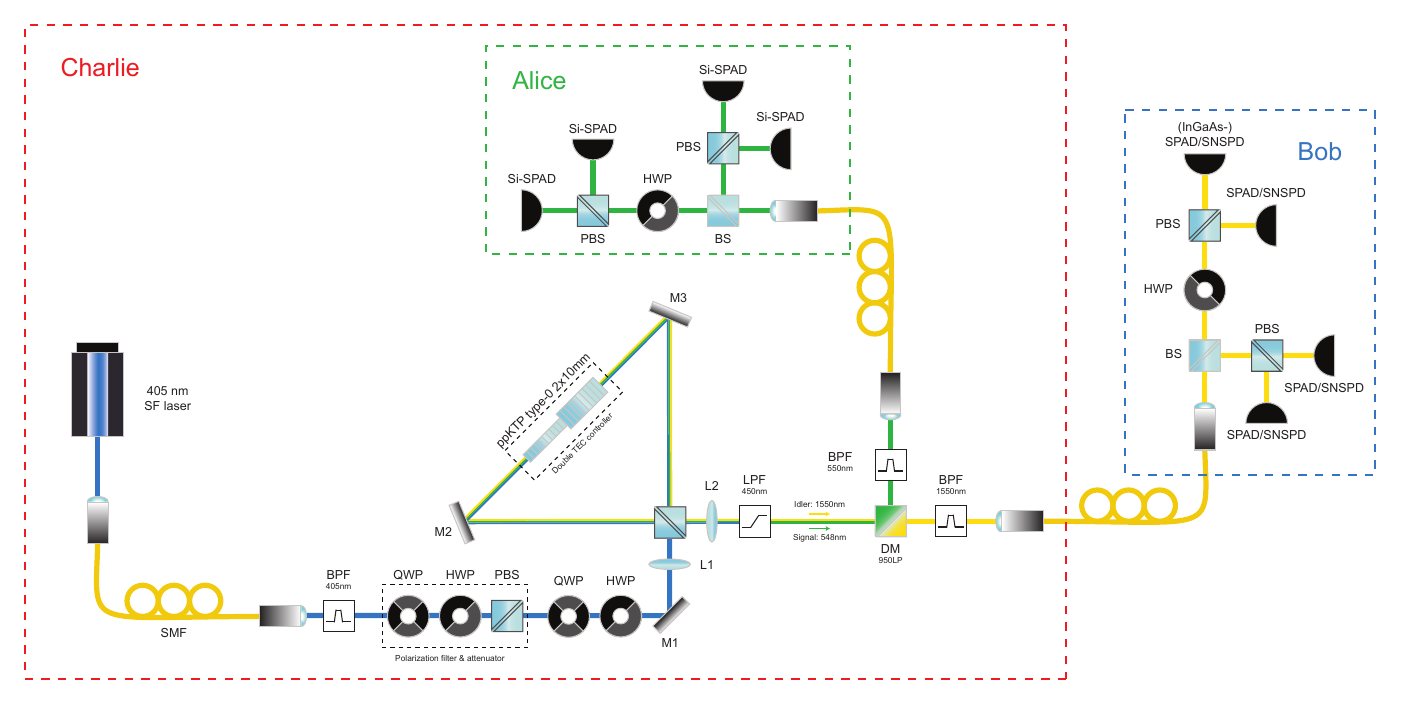}
\caption{\label{fig:scenarios}\textbf{Link scenarios D1 and D2 with the entangled configuration of the source.} The red box (Charlie) is the source node. A 405~nm single-frequency (SF) laser pumps two 10~mm type-0 ppKTP crystals inside a Sagnac loop formed by a polarising beamsplitter (PBS) and the mirrors M2 and M3. The QWP and HWP before mirror M1 set the pump polarisation. A long-pass filter (LPF, 450~nm) removes the pump, and a dichroic mirror (DM, 950~nm long-pass) separates the 548~nm signal from the 1550~nm idler. Alice (green dashed box) sits at the source node and detects the 548~nm photon with four Si-SPADs behind a passive basis choice: a beamsplitter (BS) and two PBSs, one of them after a half-wave plate (HWP). The 1550~nm photon travels through fibre to Bob (blue box), who uses the same analyser with a four-channel SNSPD (D1) or four InGaAs SPADs (D2). The two scenarios differ only in Bob's detectors.}

\end{figure*}

For scenarios D1 and D2 (Fig.~\ref{fig:scenarios}) we assume a polarisation-entangled type-0 SPDC source, in which two ppKTP crystals sit inside a Sagnac interferometer. The state of this source has the form
\begin{equation}\label{source1}
\ket{\Psi} = \sum_{j,k=0}^{\infty}\sqrt{\frac{\mu_H^j \mu_V^k}{(1+\mu_H)^{j+1}(1+\mu_V)^{k+1}}}e^{ik\varphi}\ket{j_H,k_V}_h\ket{j_H,k_V}_s,
\end{equation}
where the subscripts $h,s$ denote the modes of the two photons of a pair, $H,V$ denote the polarisations of the photons, $\varphi$ is the relative phase set by the Sagnac loop and $\mu_H,\mu_V$ are the mean numbers of H- and V-polarised pairs in one temporal mode. By setting $\mu_H=\mu_V=\mu/2$ and $n=j+k$ we obtain the balanced source
\begin{equation}
\ket{\Psi} = \frac{1}{1+\mu /2}\sum_{n=0}^{\infty}\sqrt{n+1}\left(\frac{\mu/2}{1+\mu/2}\right)^{n/2}\ket{\Phi_n},
\end{equation}
where $\mu=\mu_H+\mu_V$ is the total mean pair number per mode, and the $n$-pair polarisation state is given by

\begin{align}
    \ket{\Phi_n} = \frac{1}{\sqrt{n+1}}\sum_{k=0}^{n}e^{ik\varphi}\ket{(n-k)_{H},k_V}_h \ket{(n-k)_{H},k_V}_s.
\end{align}
The thermal statistics of Eq.~\eqref{source1} hold for one temporal mode. With a continuous pump, one coincidence window holds many temporal modes, so the number of pairs per window is close to Poissonian, with mean $R_{\rm pair}\tau_c$. At the pair numbers of this paper, below 0.05 per window, the two distributions give nearly the same coincidence rate. The rate model below therefore uses only the mean pair rate.

\begin{table*}[htb!]
\caption{\label{tab:scenarios}Link scenarios, labelled by source family (D: highly nondegenerate; G: generic comparator source). In every row the source sits at one node, so one arm is detected locally with no channel loss and only the remote arm travels. The cryostat count is the number of SNSPD systems the link needs.}
\begin{ruledtabular}
\begin{tabular}{l l l l l c}
 & Source & Local arm & Remote arm & Protocol & Cryo. \\
\colrule
\multicolumn{6}{l}{\textit{Highly nondegenerate (this source)}}\\
D1 & 548/1550 EPS & Si-SPAD 548 & SNSPD 1550 & BBM92 & 1 \\
D2 & 548/1550 EPS & Si-SPAD 548 & InGaAs 1550 & BBM92 & 0 \\
\colrule
\multicolumn{6}{l}{\textit{Generic comparator source}}\\
G1 & $2\times$1550 EPS & SNSPD 1550 & SNSPD 1550 & BBM92 & 2 \\
G3 & $2\times$1550 EPS & SNSPD 1550 & InGaAs 1550 & BBM92 & 1 \\
G2 & $2\times$1550 EPS & InGaAs 1550 & InGaAs 1550 & BBM92 & 0 \\
\end{tabular}
\end{ruledtabular}
\end{table*}

\subsection{Key-rate model}
We use a rate-based asymptotic model. It reduces each scenario to a per-event secret-key yield and a pair-emission rate. Every quantity below carries an arm index, so no step assumes that Alice and Bob hold the same detector. Distances refer to the end-to-end fibre length $L$; the 1550~nm channel transmittance over standard single-mode fibre (SMF-28, with the conventional loss coefficient $\alpha = 0.2$~dB/km~\cite{pirandola2020advances}) is
\begin{equation}
t_{\rm ch}(L) = 10^{-\alpha L/10}.
\label{eq:channel}
\end{equation}
A 548~nm photon is detected at the source node and does not travel (Rayleigh scattering alone makes silica fibre lose about 10~dB/km at 548~nm), so the local arm has unit channel transmittance. The per-arm efficiencies are
\begin{equation}
\eta_A = \eta^{A}_{\rm meas}\,\frac{\eta^{A}_{\rm det}}{\eta^{A}_{\rm lab}}\,\eta_{\rm loop}, \qquad
\eta_B = \eta^{B}_{\rm meas}\,\frac{\eta^{B}_{\rm det}}{\eta^{B}_{\rm lab}}\,\eta_{\rm loop}\,t_{\rm ch}(L),
\label{eq:armeff}
\end{equation}
where $\eta^{X}_{\rm meas}$ is the \emph{measured} heralding efficiency of arm $X$, corrected for the dead time of the detector on arm $X$ itself. Before this correction, it equals $C/S_{\bar X}$ of Table~\ref{tab:source}, where $\bar X$ is the other arm. The efficiency $\eta^{X}_{\rm lab}$ is that of the detector used in the measurement, and $\eta^{X}_{\rm det}$ is that of the detector deployed in the link. The factor $\eta_{\rm loop}$ is the transmission of the Sagnac loop for each photon. The loop factor applies to D1 and D2 only; it is unity for the generic links.

A heralding efficiency is a ratio of counted events and therefore already contains the detector used in the measurement. Equation~\eqref{eq:armeff} rescales it by the ratio of the deployed to the laboratory detector efficiency instead of factorising it into a detector efficiency and an optical efficiency. The ratio is unity when the deployed detector is the one used in the laboratory, which is the case for the 548~nm arm of D1 and D2. We attribute most of the remaining loss to mode matching. A loss budget of the module is not yet available.

At the brightnesses considered, the detectors approach their count-rate ceiling. We therefore apply a non-paralyzable dead-time correction. After a click, a detector is blind for a fixed dead time $\tau^{X}_{\rm dead}$~\cite{eisaman2011invited}. A photon that arrives during this time does not extend it. The dead time reduces the registered rate by the live fraction
\begin{equation}
\ell_X = \frac{1}{1 + S^{\rm inc}_X\,\tau^{X}_{\rm dead}}, \qquad
S^{\rm inc}_X = R_{\rm pair}\,\eta_X/n + d_X,
\label{eq:deadtime}
\end{equation}
\begin{equation}
S_X = n\,S^{\rm inc}_X\,\ell_X\,(1+p^{X}_{\rm ap}),
\label{eq:singles}
\end{equation}
where $R_{\rm pair}$ is the emitted pair rate, $S^{\rm inc}_X$ the incident singles rate on one detector of arm $X$ (detected pairs $R_{\rm pair}\eta_X$, shared by the $n$ detectors of the receiver, plus the dark-count rate $d_X$), $\ell_X$ the fraction of time the detector is live, $p^{X}_{\rm ap}$ the afterpulse probability per click~\cite{eisaman2011invited}, and $S_X$ the resulting registered singles rate of the whole receiver. Each receiver is a passive basis analyser with $n=4$ detectors (Fig.~\ref{fig:scenarios}), so each detector receives a quarter of the photons of its arm and has its own dead time and dark counts.

In an entanglement (BBM92) link a sifted bit comes from a coincidence between the two arms, with two contributions:
\begin{align}
C_{\rm true} &= R_{\rm pair}\,\eta_A\eta_B\,\ell_A\ell_B, \label{eq:ctrue}\\
C_{\rm acc}  &= S_A\,S_B\,\tau_c, \qquad
C = C_{\rm true} + C_{\rm acc}. \label{eq:cacc}
\end{align}
Here $C_{\rm true}$ is the rate of genuine pair coincidences (both photons of an emitted pair detected, each arm live). The rate $C_{\rm acc}$ is that of accidental coincidences, in which two uncorrelated singles fall within the coincidence window $\tau_c$. For independent counters, this rate is the product of the registered singles rates and the window~\cite{telecom_eps_2025}. Afterpulses appear only inside $S_X$, and hence only inside $C_{\rm acc}$.

The coincidence window is set by the combined timing jitter of the pair, $\tau_c = 3\sqrt{j_A^2+j_B^2}$ with $j_X$ the FWHM jitter of detector $X$, giving \SI{0.11}{ns} for two SNSPDs (G1), \SI{0.46}{ns} for an SNSPD with an InGaAs SPAD (G3), \SI{0.64}{ns} for two InGaAs SPADs (G2), \SI{0.90}{ns} for a Si-SPAD with an SNSPD (D1) and \SI{1.01}{ns} for a Si-SPAD with an InGaAs SPAD (D2). The Si-SPAD has the largest jitter, 300~ps at 300~kcps~\cite{qubitrium_qdcv}. The bench measurement implies a window of 1.2--1.6~ns (Fig.~\ref{fig:source_scaling}), somewhat wider than the 1.0~ns of the model for the same detectors. The jitter of the Si-SPAD grows with its count rate, up to 1~ns~\cite{qubitrium_qdcv}. With 1~ns, the target D1 of Sec.~\ref{sec:results} would cost \$71 instead of \$27 per Gbit. 

The model does not include chromatic dispersion in the fibre. Standard fibre spreads a 1550~nm photon by about 17~ps per nm of bandwidth and per km~\cite{telecom_eps_2025}. For the predicted bandwidth of 1.5~nm (Sec.~\ref{sec:source}), this is about 1.3~ns at 50~km and 2.6~ns at 100~km, more than the windows above. A link at these distances would need a wider window, a narrower filter or dispersion compensation. Each of these measures lowers the key rate. The model also assumes that the polarisation drift in the fibre is compensated, as in Ref.~\cite{telecom_eps_2025}. The rates at 50 and 100~km are therefore optimistic, for this source and for the generic links. In a rough estimate, each link uses its best coincidence window. Dispersion then lowers every rate at 50~km and raises every cost per bit, but the order of the links in cost per bit stays the same. In rate, only the target D2 falls slightly below G3.

True coincidences carry the intrinsic misalignment error $e_d$, whereas accidental coincidences are uncorrelated and carry the random-event error $e_0 = 1/2$, giving the quantum bit error rate (QBER)
\begin{equation}
E = \frac{e_d\,C_{\rm true} + e_0\,C_{\rm acc}}{C}.
\label{eq:qber}
\end{equation}
The asymptotic BBM92 secret-key rate follows Ma, Fung and Lo~\cite{ma2007quantum},
\begin{equation}
R_{\rm sec} = q\,C\,\big[\,1 - f\,H_2(E) - H_2(E)\,\big],
\label{eq:bbm92}
\end{equation}
where $q$ is the basis-sifting factor, $f \ge 1$ the error-correction efficiency relative to the Shannon limit, and $H_2(x) = -x\log_2 x - (1-x)\log_2(1-x)$ the binary entropy.

\subsection{Parameters}\label{sec:params}
Model parameters are collected in Table~\ref{tab:params} and detector parameters in Table~\ref{tab:detectors}. Many entries are assumptions. Two of them are choices that shape the comparison. The first is the generic $2\times$1550 source. We take both of its numbers from one published source, a type-0 PPLN crystal in a Sagnac loop~\cite{neumann2022experimental}. In one 100~GHz grid channel, with a passband of about 0.6~nm, it generates $4.1\times10^{6}$~pairs/s per mW and per nm, or $3.6\times10^{7}$~pairs/s at our pump power of 14.5~mW. Its heralding efficiency is 25.9\% with SNSPDs of 80\% efficiency, so its collection is 0.32. This is lower than the 0.52 of Ref.~\cite{telecom_eps_2025}, because a tight pump focus raises the brightness but lowers the coupling into fibre (Sec.~\ref{sec:source}). All links run at the same pump power, and Sec.~\ref{sec:results} gives the effect of a different pump power. The second is the \SI{100}{b/s} floor $R_{\min}$. It is an operational threshold, roughly one AES-256 rekey every few seconds. The underlying rate is listed either way.

Table~\ref{tab:params} gives two parameter sets. The \emph{current} set describes the module as measured in Sec.~\ref{sec:source}. The \emph{target} set describes an improved module. It changes three inputs of this source: the collection, the brightness and the loop transmission. The target collection is our estimate for a published source (Sec.~\ref{sec:source}). The doubled brightness was measured on the optical bench, not on the module. The loop transmission is expected but not yet measured.

The misalignment error $e_d=2\%$ corresponds to a fidelity of about 97\% for white noise. This is above the target fidelity of 95\% in Sec.~\ref{sec:source}. At a fidelity of 95\%, $e_d$ would be 3.3\%, and every rate at 50~km would be about 23\% lower.

\begin{table*}[htb!]
\caption{\label{tab:params}Model parameters for the current and the target parameter sets. An empty Target cell means that the target set keeps the current value. In the target set, the heralding efficiencies are those implied by the target collection with the laboratory detectors. Detector-specific values are in Table~\ref{tab:detectors}.}
\begin{ruledtabular}
\begin{tabular}{l l l l}
Symbol & Current & Target & Meaning \\
\colrule
\multicolumn{4}{l}{\textit{Channel, protocol and this source}}\\
$\alpha$ & 0.2~dB/km & & SMF-28 loss at 1550~nm~\cite{pirandola2020advances} \\
$\eta^{548}_{\rm meas}$ & 0.111 & 0.43 & 548~nm heralding efficiency, $C/S_2$ (Table~\ref{tab:source}), dead-time corrected; raw 9.7\% \\
$\eta^{1550}_{\rm meas}$ & 0.046 & 0.10 & 1550~nm heralding efficiency, $C/S_1$, dead-time corrected; raw 1.6\% \\
$\eta^{548}_{\rm lab}$ & 0.82 & & Si-SPAD efficiency during that measurement (Sec.~\ref{sec:design}) \\
$\eta^{1550}_{\rm lab}$ & 0.20 & & IDQ InGaAs efficiency during that measurement \\
$\eta_{\rm coll}$ & 0.135 / 0.230 & 0.52 / 0.52 & collection efficiency of the two arms, $\eta_{\rm meas}/\eta_{\rm lab}$; target from Ref.~\cite{telecom_eps_2025} \\
$n$ & 4 & & detectors per receiver, passive basis choice (Fig.~\ref{fig:scenarios}) \\
$\eta_{\rm loop}$ & 0.85 & 0.90 & Sagnac transmission per photon, entangled configuration (D1, D2) \\
$e_d$ & 0.02 & & intrinsic misalignment error \\
$e_0$ & 1/2 & & error rate of an uncorrelated coincidence \\
$f$ & 1.16 & & error-correction efficiency~\cite{lim2014concise} \\
$q$ & 1/2 & & basis-sifting factor \\
$\tau_c$ & $3\sqrt{j_A^2+j_B^2}$ & & 0.11 to 1.01~ns (see text) \\
$R_{\min}$ & 100~b/s & & operational floor for a usable link \\
\colrule
\multicolumn{4}{l}{\textit{Sources}}\\
$R_{\rm pair}$ & $2.11\times10^{7}$~s$^{-1}$ & $4.23\times10^{7}$~s$^{-1}$ & 548/1550 EPS (D1, D2), $S_1S_2/C$ at 14.5~mW; target: single-mode pump \\
$R_{\rm pair}$ & $3.6\times10^{7}$~s$^{-1}$ & & generic $2\times$1550 EPS (G1, G3, G2), 100~GHz channel at 14.5~mW~\cite{neumann2022experimental} \\
$\eta_{\rm meas}$ & 0.259 & & generic $2\times$1550 EPS arm, heralding efficiency~\cite{neumann2022experimental} \\
$\eta_{\rm lab}$ & 0.80 & & detector efficiency in that measurement \\
\end{tabular}
\end{ruledtabular}
\end{table*}

All values are asymptotic. Finite-key penalties, which we do not model, reduce every rate and matter most for the lowest-rate links at long distance~\cite{tomamichel2012tight,lim2014concise}. For BBM92, only the coincidence rate and the QBER have to be estimated from the data~\cite{ma2007quantum}.

\begin{table}[htb!]
\caption{\label{tab:detectors}Representative detector parameters. $p_{\rm ap}$ is the afterpulse probability per click, which enters only through the singles rate [Eq.~\eqref{eq:singles}]. Provenance: Si, PDE measured in house at 548~nm, the remaining values from the datasheet of the Qubitrium QDC-V2 module~\cite{qubitrium_qdcv}, with the jitter at 300~kcps and $p_{\rm ap}$ at its upper bound of 0.1\%; InGaAs, 20\% PDE with the hold-off and afterpulsing tied by the power law of Ref.~\cite{xu2024freerunning}, which gives 2.7\% at a 5~$\mu$s hold-off; this hold-off is longer than the 2~$\mu$s of the measurement, so the InGaAs rates are conservative; SNSPD, a top-tier commercial system quoting 1--100~cps over 1300--1600~nm~\cite{idq_id281}, with the PDE at the top of its specified range, against which laboratory devices reach \num{98.0(5)}\%~\cite{reddy2020superconducting} and $99.5^{+0.5}_{-2.1}$\% at 1350~nm~\cite{chang2021detecting}.}
\begin{ruledtabular}
\begin{tabular}{l c c c c c c}
Detector & $\lambda$ & PDE & dark & jitter & $\tau_{\rm dead}$ & $p_{\rm ap}$ \\
 & (nm) & & (cps) & (ps) & & \\
\colrule
Si-SPAD      & 548  & 0.82 & 100  & 300 & 60~ns & 0.1\% \\
InGaAs-SPAD  & 1550 & 0.20 & 1000 & 150 & 5~$\mu$s & 2.7\% \\
SNSPD        & 1550 & 0.95 & 10   & 25  & 30~ns & 0\% \\
\end{tabular}
\end{ruledtabular}
\end{table}

\subsection{Results}\label{sec:results}

Table~\ref{tab:results} gives the secret-key rate, the QBER at 0~km and the per-emission efficiency $R_{\rm norm}$ (secret bits per emitted pair) for both parameter sets. In the current set, G1 leads on every measure of performance. The other links trade performance for fewer cryostats and lower capital. The rows for this source, D1 and D2, use the efficiencies of the unoptimised module of Sec.~\ref{sec:source}. They describe its present state and not the potential of the design. The points below refer to the current set. The target set follows after them.
\begin{itemize}

\item \emph{Throughput and per-emission efficiency:} the generic link G1 leads in both, at every distance. It reaches $9.7\times10^{5}$~b/s at 0~km, about $9\times$ D1, and $2.7\times10^{-2}$ bits per emitted pair against D1's $4.9\times10^{-3}$. With four detectors per receiver, the SNSPDs of G1 are live 92\% of the time at 0~km, so detector saturation limits this link only weakly.
\item \emph{Detector saturation with low-cost detectors:} pairing the bright source with free-running InGaAs (G2) gives $\ell\!\approx\!0.26$. The rate collapses to about 2900~b/s and flattens with distance, because the remote arm leaves saturation only when the channel loss grows. The 5~$\mu$s hold-off causes this saturation. D2 makes 2.3 times more key than G2 at 50~km.
\item \emph{The one-cryostat comparison:} G3 is the fair competitor to D1, since both need a single SNSPD. D1 leads up to about 26~km, with $1.0\times10^{5}$ against $5.2\times10^{4}$~b/s at 0~km, because the bright source of G3 saturates its InGaAs detectors on the travelling arm, which are live only 26\% of the time. Beyond about 26~km, G3 delivers more key. Its saturated rate falls slowly with distance, while the rate of D1 falls with the channel loss from the start: $1.6\times10^{4}$ against $1.1\times10^{4}$~b/s at 50~km. G3 is also cheaper per bit at 50~km, \$301 against \$376 per Gbit. With the present module, D1 is therefore the better one-cryostat link up to about 26~km, and G3 beyond.

\item \emph{QBER:} every link sits above the misalignment floor $e_d = 2\%$. The excess comes from accidental coincidences. Per true coincidence, their number is $R_{\rm pair}\tau_c(1+p^{A}_{\rm ap})(1+p^{B}_{\rm ap})$, because the live fractions cancel between Eqs.~\eqref{eq:ctrue} and \eqref{eq:cacc}. The QBER therefore follows the number of pairs per coincidence window, which grows with the brightness and with the detector jitter. The order is G1 $2.18\%$, G3 $2.79\%$, D1 $2.90\%$, D2 $3.03\%$ and G2 $3.13\%$. The jitter of the Si-SPAD places D1 and D2 above G3. Afterpulsing adds less than 0.1 percentage points, and dead time does not change the QBER.
\item \emph{Pump power:} all links run at the pump power of our module, 14.5~mW. Each link has its own best pump power. G1 gains most from a stronger pump, because its SNSPDs are fast and its coincidence window is short. At 50~km and about 13 times the pump power, G1 makes $4.5\times10^{5}$~b/s and costs \$15 per Gbit. D1 does best at about 4.5 times the pump power, at \$166 per Gbit, and G3 at about 2.8 times, at \$217. We measured our module only up to 14.8~mW.
\end{itemize}

The collection efficiency drives the D rows of the table. The module collects $0.135$ of the pairs it makes on the visible arm and $0.230$ on the telecom arm (Sec.~\ref{sec:source}), and every D row uses both values. The module was not optimised for collection, so these values are a starting point. Brightness does not explain the gap, because $R_{\rm norm}$ is already normalised per emitted pair. The pairs are generated and then lost before they reach a detector. These collection losses are not a fundamental limit. They define a design axis and not a ceiling. Section~\ref{sec:collection} analyses this axis.

Apart from collection, the design offers four features: a single-frequency 405~nm pump laser, a visible arm at 548~nm where the silicon detector runs at $82\%$, close to its peak, a miniaturised module instead of an optical bench, and, once the entangled configuration works, a link with one cryostat and a local node that needs only Peltier cooling.

\emph{Target parameter set.} The second block of Table~\ref{tab:results} shows the improved module. The generic rows do not change, because the target set improves only this source. At 50~km, D1 rises about fourteenfold, from $1.1\times10^{4}$ to $1.5\times10^{5}$~b/s, and D2 rises about tenfold, from $2.0\times10^{3}$ to $2.0\times10^{4}$~b/s. The QBER is 3.8\% for D1 and 4.0\% for D2. Collection gives most of the gain: on its own it raises D1 about eightfold and D2 about sevenfold. Doubling the brightness adds about 55\% to D1 and 30\% to D2.

With these values, D1 overtakes the two-cryostat link G1. It costs \$27 per Gbit over two years against \$66 for G1, and it makes about 1.4 times more key at every distance, with one cryostat instead of two (Sec.~\ref{sec:design}). D1 also leads G3 at every distance and costs about a tenth as much per bit. D2 needs no cryostat at all and costs \$90 per Gbit, about 70\% less than G3 (\$301), and it makes about 1.3 times more key than G3 at 50~km. G1 keeps the lowest QBER. If every link runs at its own best pump power, G1 is again the cheapest: \$15 per Gbit against \$23 for the target D1.

A faster InGaAs setting would raise D2 further. At a 1~$\mu$s dead time, the setting of our second measured run in Fig.~\ref{fig:source_scaling}, D2 would rise by about 27\% to $2.6\times10^{4}$~b/s at 50~km, with an afterpulse probability of 22\% extrapolated from Ref.~\cite{xu2024freerunning}. The same setting would also help the generic rows, G3 by a factor of 1.2 and G2 by a factor of 2.8, so we keep the detectors the same in both sets. With the faster setting in every row, D2 would still cost about 70\% less per bit than G3, \$71 against \$252 per Gbit.

\begin{table*}[t]
\caption{\label{tab:results}Results for the current and the target parameter sets of Table~\ref{tab:params}, grouped by source family. $R_{\rm sec}$ in bits/s, with the end-to-end fibre length in km in parentheses. $R_{\rm norm}$ is the secret bits per emitted pair at the source node, $R_{\rm sec}(0)/R_{\rm pair}$. SMF-28 at 0.2~dB/km. The QBER is given at 0~km. Cryo.\ is the number of cryostats.}
\small
\begin{ruledtabular}
\begin{tabular}{l l c c c c c c}
Scenario & Detectors & Cryo. & QBER & $R_{\rm norm}$(0) & $R_{\rm sec}$(0) & $R_{\rm sec}$(50) & $R_{\rm sec}$(100) \\
\colrule
\multicolumn{8}{l}{\textbf{Current parameter set}}\\
\multicolumn{8}{l}{\textit{Highly nondegenerate (this source)}}\\
D1 & Si/SNSPD       & 1 & 2.90\% & $4.9\times10^{-3}$ & $1.0\times10^{5}$ & $1.1\times10^{4}$ & $1.1\times10^{3}$ \\
D2 & Si/InGaAs      & 0 & 3.03\% & $5.2\times10^{-4}$ & $1.1\times10^{4}$ & $2.0\times10^{3}$ & 201 \\
\colrule
\multicolumn{8}{l}{\textit{Generic comparator source}}\\
G1 & SNSPD/SNSPD  & 2 & 2.18\% & $2.7\times10^{-2}$ & $9.7\times10^{5}$ & $1.0\times10^{5}$ & $1.1\times10^{4}$ \\
G3 & SNSPD/InGaAs & 1 & 2.79\% & $1.4\times10^{-3}$ & $5.2\times10^{4}$ & $1.6\times10^{4}$ & $1.9\times10^{3}$ \\
G2 & InGaAs/InGaAs& 0 & 3.13\% & $8.0\times10^{-5}$ & $2.9\times10^{3}$ & 859 & 104 \\
\colrule
\multicolumn{8}{l}{\textbf{Target parameter set}}\\
\multicolumn{8}{l}{\textit{Highly nondegenerate (this source)}}\\
D1 & Si/SNSPD & 1 & 3.77\% & $3.1\times10^{-2}$ & $1.3\times10^{6}$ & $1.5\times10^{5}$ & $1.5\times10^{4}$ \\
D2 & Si/InGaAs & 0 & 4.01\% & $1.2\times10^{-3}$ & $5.1\times10^{4}$ & $2.0\times10^{4}$ & $2.8\times10^{3}$ \\
\colrule
\multicolumn{8}{l}{\textit{Generic comparator source: the same as in the current set}}\\
\end{tabular}
\end{ruledtabular}
\end{table*}

\subsection{Cost}

Table~\ref{tab:cost} gives the hardware capital of each link and its cost per secret bit. A link contains one pair source and one receiver per arm, as in the rate model. A receiver with SPADs has four detectors, one for each output of the passive basis analyser. An SNSPD receiver is one four-channel system with one cryostat. We take one price per component: SNSPD system \$200~k, free-running InGaAs SPAD \$15~k, Si-SPAD \$5~k and pair source \$35~k. The Si-SPAD price follows the distributor list prices of a commercial fibre-coupled module (Excelitas SPCM-AQRH). The other three are order-of-magnitude estimates, because we found no public list prices for them. The cost per bit is the capital divided by the key that the link makes in one or two years of continuous operation at its 50~km rate. Fibre, electronics and operating cost are not included.

The calculation has two steps. First, the capital of a link is the sum of its parts, in k\$: D1 $=35+4\times5+200=255$, D2 $=35+4\times5+4\times15=115$, G1 $=35+2\times200=435$, G3 $=35+200+4\times15=295$ and G2 $=35+2\times4\times15=155$. Second, the cost per secret bit divides this capital by the key that the link makes in $T$ years, $c_{\rm bit}={\rm CAPEX}/[R_{\rm sec}(50)\,T]$, with one year equal to $3.156\times10^{7}$~s. For example, D1 in the current set makes $1.07\times10^{4}$~b/s, or $3.39\times10^{11}$ bits per year, and its capital of 255~k\$ gives \$752 per Gbit over one year and \$376 over two.

\begin{table*}[t]
\caption{\label{tab:cost}Hardware capital and cost per secret bit per point-to-point link, amortising capital over one or two years of continuous operation at the 50~km rate $R_{\rm sec}(50)$ of Table~\ref{tab:results}, for the current and the target parameter sets. Prices and what they include are given in the text.}
\small
\begin{ruledtabular}
\begin{tabular}{l c c c c c}
Scenario & CAPEX (k\$) & Cryo. & $R_{\rm sec}(50)$ [b/s] & \$/Gbit (1\,yr) & \$/Gbit (2\,yr) \\
\colrule
\multicolumn{6}{l}{\textbf{Current parameter set}}\\
D1 & 255 & 1 &  $1.1\times10^{4}$ &  752 &  376 \\
D2 & 115 & 0 &  $2.0\times10^{3}$ &  $1.8\times10^{3}$ &  917 \\
G1 & 435 & 2 &  $1.0\times10^{5}$ &  132 &  66 \\
G3 & 295 & 1 & $1.6\times10^{4}$ & 602 & 301 \\
G2 & 155 & 0 &  859 &  $5.7\times10^{3}$ &  $2.9\times10^{3}$ \\
\colrule
\multicolumn{6}{l}{\textbf{Target parameter set}}\\
D1 & 255 & 1 & $1.5\times10^{5}$ & 54 & 27 \\
D2 & 115 & 0 & $2.0\times10^{4}$ & 181 & 90 \\
\multicolumn{6}{l}{\textit{G1, G3, G2: the same as in the current set}}\\
\end{tabular}
\end{ruledtabular}
\end{table*}

In the current set, G1 is the cheapest per bit, at \$66 per Gbit over two years. Its high rate spreads the capital of two cryostats. G3 follows at \$301, and D1 costs \$376. In the target set, D1 falls to \$27, less than half the cost of G1, with one cryostat instead of two, and D2 falls to \$90 with no cryostat at all, well below G3. G2 stays the most expensive link, at \$2\,900 per Gbit, because its bright source saturates the slow InGaAs detectors.

\subsection{Collection efficiency sets the design point}\label{sec:collection}

\begin{figure*}[t]
\includegraphics[width=\textwidth]{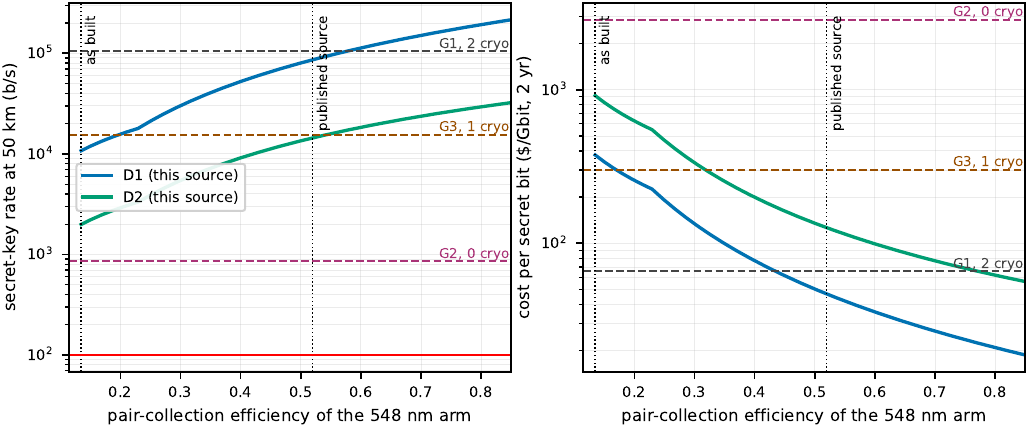}
\caption{\label{fig:collection}\textbf{Collection efficiency as the design axis.} Secret-key rate at 50~km (left) and cost per secret bit over two years (right) against the pair-collection efficiency of the 548~nm arm, for the two link designs built on this source. The telecom arm is held at its measured $0.230$ until the visible arm overtakes it and follows it afterwards. This causes the kink near $0.23$. The generic comparators G1, G3 and G2 are independent of this axis and appear as horizontal lines, labelled by cryostat count. Dotted verticals mark the unoptimised module as built ($0.135$) and the collection efficiency that we estimate for the published nondegenerate source of Ref.~\cite{telecom_eps_2025} ($\approx0.52$ on both arms; Sec.~\ref{sec:source}). This is also the collection of the target set (Table~\ref{tab:params}), which in addition doubles the brightness and raises the loop transmission. The red line in the left panel is the 100~b/s operational floor.}
\end{figure*}

The two arms have unequal collection efficiencies: $0.135$ at 548~nm and $0.230$ at 1550~nm. The visible arm limits the link, and it has the most room for improvement. In Fig.~\ref{fig:collection} we sweep its collection efficiency. The telecom arm follows once the visible arm overtakes it. Nothing else in the model moves. The horizontal axis is therefore one quantity that the bench can change: how much of the emitted light reaches a fibre. The generic rows do not move with it and appear as fixed lines. The sweep uses the current parameter set.

Figure~\ref{fig:collection} shows where each comparison changes. In cost per bit, D1 matches the one-cryostat link G3 at a visible collection of $0.17$, about 1.3 times the built value. D2 matches G3 at $0.32$, about 2.4 times the built value. At this point, a link without a cryostat costs the same as a link with one. D1 matches the two-cryostat link G1 at $0.43$. At $0.52$, the value that we estimate for Ref.~\cite{telecom_eps_2025}, D1 reaches $8.6\times10^{4}$~b/s and \$47 per Gbit. The target set combines this collection with the doubled brightness and the better loop. This brings D1 to \$27 per Gbit, less than half the cost of G1 (Table~\ref{tab:cost}).

 Raising the visible collection from $0.135$ to $0.25$, and with it the telecom collection from $0.230$ to $0.25$, about halves the cost per bit of D1, from \$376 to \$191 per Gbit. This gain depends on mode matching and fibre coupling, not on the detectors. The present module is not optimised for it.

\section{Security analysis}\label{sec:security}

\paragraph*{Entanglement is the precondition.} BBM92~\cite{bennett1992quantum} draws its security from the entanglement between the two photons. The module measured in Sec.~\ref{sec:source} is expected to produce a separable mixture: its pairs are correlated in the H/V basis but not in the diagonal basis. A BBM92 link built on it would see an error rate of 50\% in the diagonal basis and would yield no secret key. The key rates of Sec.~\ref{sec:design} therefore assume the entangled configuration with the error rate $e_d=2\%$ of Table~\ref{tab:params}. Every statement below depends on this configuration. The heralding efficiencies and the CAR are already measured on the correlated module (Table~\ref{tab:source}). What remains is to measure the entangled configuration itself: its visibility in both bases, or its fidelity, at the operating pump power.

\paragraph*{Multi-pair emission.} An entangled down-conversion source is basis-independent, so the key-rate formula of Eq.~\eqref{eq:bbm92} holds for its full multi-pair statistics~\cite{ma2007quantum}. BBM92 therefore has no counterpart of the photon-number-splitting attack that limits weak-coherent-pulse BB84 without decoy states~\cite{brassard2000limitations}. Multi-pair emission instead adds accidental coincidences and raises the QBER, which Eq.~\eqref{eq:qber} includes. The rates are asymptotic, and finite-key effects reduce them (Sec.~\ref{sec:design}).

\paragraph*{Detector side channels.} Detector side channels are the main practical risk for every link compared here. The detectors sit inside the trusted nodes, and Bob's receiver at the far end of the fibre is open to an eavesdropper through the channel. Bright light can take control of a detector. This was shown for the gated InGaAs avalanche photodiodes of two commercial systems~\cite{lydersen2010hacking} and for an SNSPD in the laboratory~\cite{lydersen2011controlling}. In a complete attack on a running BBM92 link with polarisation-entangled photons and Si avalanche photodiodes, the eavesdropper obtained the full key without raising the QBER~\cite{gerhardt2011full}. Time-shift attacks exploit detectors in one receiver whose efficiencies differ in time~\cite{Qi_2007}, which matters for the four-detector passive receivers of Fig.~\ref{fig:scenarios}. The detector-matched design neither adds nor removes these attack classes, because every detector type, in this design and in the generic links, is open to them. Countermeasures exist for each attack. Measurement-device-independent QKD removes detector side channels altogether~\cite{lo2012measurement}, but it needs two-photon interference between independent sources and lies outside the scope of this paper.

\paragraph*{Dead-time attack.} A long detector dead time opens a further attack. The eavesdropper sends dim pulses of one polarisation into the fibre. They make some of Bob's detectors click, and these detectors are still dead when the next photon arrives. That photon can then reach only the remaining detectors, so its bit becomes predictable. In a BB84 experiment with a four-detector receiver, this attack revealed up to 98.8\% of the key without a significant rise in the QBER~\cite{weier2011quantum}. The countermeasure keeps only the events in which all of Bob's detectors are live~\cite{weier2011quantum}. Our model gives each detector its own dead time and does not include this countermeasure. With it, the receivers with InGaAs detectors lose key, because their dead time is long. At 50~km, G3 would cost \$506 instead of \$301 per Gbit, and D2 \$1185 instead of \$917. D1 and G1 change by less than 5\%, so D1, at \$380, would then cost less per bit than G3. In the target set, D2 would cost \$181 instead of \$90, still about 65\% less than G3 with the same countermeasure (\$506).

\paragraph*{Comparison with decoy BB84.} Decoy-state BB84 with weak coherent pulses is the mature and deployed baseline~\cite{xu2020secure,pirandola2020advances}. In a simulation with the parameters of a 144~km free-space experiment, it reached higher key rates than an entangled source at low and medium loss~\cite{ma2007quantum}. Its transmitter must be trusted and well characterised, including against Trojan-horse probing~\cite{navarrete2022practical} and correlations between pulse intensities~\cite{zapatero2021security}. A BBM92 source holds no secret bit or basis value and needs no active modulation. The key bits come from measurement outcomes. With passive basis choice, a beamsplitter sets the basis, so no random-number generator is needed. An eavesdropper can reach Bob's beamsplitter through the fibre and control his basis choice with light at other wavelengths~\cite{li2011attacking}, so this choice must be protected. In the links studied here, however, the source sits at Alice's node, so we do not use the option of placing an untrusted source between the users~\cite{waks2002security}. None of these advantages applies until the source is shown to be entangled.

\section{Conclusions}\label{sec:conclusions}
We have characterised a first module of a highly nondegenerate photon-pair source at 548 and 1550~nm, in its polarisation-correlated configuration. The module generates about $1.5\times10^{6}$ pairs/s per mW of pump power. Once the detector efficiencies are divided out, it collects 13\% of the pairs on the 548~nm arm and 23\% on the 1550~nm arm. These values describe a module that was not optimised for collection.

In the link model with four detectors per receiver, a link with this source and one SNSPD (D1) costs about \$376 per secret Gbit over two years at 50~km. The generic links use a bright published source at the same pump power. The generic link with one SNSPD (G3) costs \$301, and the generic link with two SNSPDs (G1) costs \$66. With the present module, D1 makes more key than G3 up to about 26~km. At its own best pump power, G1 costs about \$15 per Gbit.

A target module with 52\% collection on both arms, twice the brightness and a better loop changes this comparison. D1 then costs \$27 per Gbit, about 40\% of the cost of G1, with one cryostat instead of two, and about a tenth of the cost of G3. The link with this source and InGaAs SPADs (D2) needs no cryostat and costs \$90 per Gbit, about 70\% less than G3. At the best pump power of each link, G1 stays cheaper than the target D1, at \$15 against \$23 per Gbit. These target values are optimistic. The closest published source reaches an estimated collection of about 52\%, but its brightness is 0.56~MHz/mW, about a fifth of the 2.9~MHz/mW of the target set~\cite{telecom_eps_2025}. Most of the gain from the current to the target set comes from collection (Sec.~\ref{sec:results}).

These results are preliminary and rest on four assumptions. First, all key rates assume the entangled configuration. The measured module is expected to emit a separable mixture and would yield no secret key. Second, the rates are asymptotic, and the model leaves out fibre dispersion and polarisation drift, which lower the rates at 50 and 100~km. Third, three of the four component prices are order-of-magnitude estimates. Fourth, all links run at the same pump power of 14.5~mW, and a two-SNSPD link gains most from a stronger pump.

Three steps come next. First, the entangled configuration has to show its fidelity, or its visibility in both bases, at the operating pump power. Second, a pump in a single spatial mode and better mode matching should raise the collection, which raises the key rate more than any other change considered here. Third, a measured afterpulse probability of the InGaAs detector would allow the heralding efficiencies to be corrected for afterpulsing.

The source also has uses beyond QKD. Clock synchronisation with photon pairs~\cite{lee2019symmetrical} and detector calibration with correlated photons~\cite{polyakov2007high} need only the time correlation of the pairs, so the present module could serve them. A QKD link built on this source could also take timing from its own coincidences, with no extra hardware~\cite{krelina2026quantum}. The precision of this timing would be limited by the 300~ps jitter of the Si-SPAD and by the fibre dispersion of the 1550~nm photon. Imaging with undetected photons~\cite{lemos2014quantum} would need a different optical layout, and a quantum-memory interface~\cite{lagorivera2021telecom} would need the visible photon at a memory transition.


\section*{References}
\bibliography{refs}

@article{ekert1991quantum,
  title={Quantum cryptography based on Bell’s theorem},
  author={Ekert, Artur K},
  journal={Physical review letters},
  volume={67},
  number={6},
  pages={661},
  year={1991},
  publisher={APS}, doi={10.1103/PhysRevLett.67.661}
}

@article{waks2002security,
  title={Security of quantum key distribution with entangled photons against individual attacks},
  author={Waks, Edo and Zeevi, Assaf and Yamamoto, Yoshihisa},
  journal={Physical Review A},
  volume={65},
  number={5},
  pages={052310},
  year={2002},
  publisher={APS}, doi={10.1103/PhysRevA.65.052310}
}

@article{Qi_2007, title={Time-shift attack in practical quantum cryptosystems}, volume={7}, ISSN={1533-7146},  number={1-2}, journal={Quantum Info. Comput.}, DOI={10.26421/QIC7.1-2-3}, author={Qi, Bing and Fung, Chi-Hang Fred and Lo, Hoi-Kwong and Ma, Xiongfeng}, year={2007}, month=jan, pages={73–82} }

@article{ma2007quantum,
  title={Quantum key distribution with entangled photon sources},
  author={Ma, Xiongfeng and Fung, Chi-Hang Fred and Lo, Hoi-Kwong},
  journal={Physical Review A}, volume={76}, number={1}, pages={012307},
  year={2007}, publisher={APS}, doi={10.1103/PhysRevA.76.012307}, note={arXiv:quant-ph/0703122}
}

@article{lim2014concise,
  title={Concise security bounds for practical decoy-state quantum key distribution},
  author={Lim, Charles Ci Wen and Curty, Marcos and Walenta, Nino and Xu, Feihu and Zbinden, Hugo},
  journal={Physical Review A}, volume={89}, number={2}, pages={022307},
  year={2014}, publisher={APS}, doi={10.1103/PhysRevA.89.022307}
}

@article{brassard2000limitations,
  title={Limitations on practical quantum cryptography},
  author={Brassard, Gilles and L{\"u}tkenhaus, Norbert and Mor, Tal and Sanders, Barry C},
  journal={Physical Review Letters}, volume={85}, number={6}, pages={1330},
  year={2000}, publisher={APS}, doi={10.1103/PhysRevLett.85.1330}
}

@article{lydersen2010hacking,
  title={Hacking commercial quantum cryptography systems by tailored bright illumination},
  author={Lydersen, Lars and Wiechers, Carlos and Wittmann, Christoffer and Elser, Dominique and Skaar, Johannes and Makarov, Vadim},
  journal={Nature Photonics}, volume={4}, number={10}, pages={686--689},
  year={2010}, publisher={Nature Publishing Group}, doi={10.1038/nphoton.2010.214}
}

@article{pirandola2020advances,
  title={Advances in quantum cryptography},
  author={Pirandola, Stefano and Andersen, Ulrik L and Banchi, Leonardo and Berta, Mario and Bunandar, Darius and Colbeck, Roger and Englund, Dirk and Gehring, Tobias and Lupo, Cosmo and Ottaviani, Carlo and others},
  journal={Advances in Optics and Photonics}, volume={12}, number={4}, pages={1012--1236},
  year={2020}, publisher={Optica}, doi={10.1364/AOP.361502}
}

@article{reddy2020superconducting,
  title={Superconducting nanowire single-photon detectors with 98\% system detection efficiency at 1550 nm},
  author={Reddy, Dileep V and Nerem, Robert R and Nam, Sae Woo and Mirin, Richard P and Verma, Varun B},
  journal={Optica}, volume={7}, number={12}, pages={1649--1653},
  year={2020}, publisher={Optica}, doi={10.1364/OPTICA.400751}
}

@article{zapatero2021security,
  title={Security of quantum key distribution with intensity correlations},
  author={Zapatero, V{\'\i}ctor and Navarrete, {\'A}lvaro and Tamaki, Kiyoshi and Curty, Marcos},
  journal={Quantum}, volume={5}, pages={602}, year={2021}, doi={10.22331/q-2021-12-07-602}
}

@article{navarrete2022practical,
  title={Improved finite-key security analysis of quantum key distribution against {Trojan}-horse attacks},
  author={Navarrete, {\'A}lvaro and Curty, Marcos},
  journal={Quantum Science and Technology}, volume={7}, number={3}, pages={035021},
  year={2022}, publisher={IOP}, doi={10.1088/2058-9565/ac74dc}
}

@misc{telecom_eps_2025,
  title={Highly nondegenerate entangled photon source for fiber-based quantum key distribution},
  author={Dosan, Vasile-Laurentiu and Lagarrigue, Alek and Zannotti, Alessandro and Parab, Tushar and Folwill, Yannick and Steinlechner, Fabian and de Vries, Oliver},
  year={2025}, note={arXiv:2512.11630},
  doi={10.48550/arXiv.2512.11630}
}

@article{lo2012measurement,
  title={Measurement-Device-Independent Quantum Key Distribution},
  author={Lo, Hoi-Kwong and Curty, Marcos and Qi, Bing},
  journal={Physical Review Letters}, volume={108}, number={13}, pages={130503},
  year={2012}, publisher={American Physical Society},
  doi={10.1103/PhysRevLett.108.130503}, note={arXiv:1109.1473}
}

@article{eisaman2011invited,
  title={Invited Review Article: Single-photon sources and detectors},
  author={Eisaman, M. D. and Fan, J. and Migdall, A. and Polyakov, S. V.},
  journal={Review of Scientific Instruments}, volume={82}, number={7}, pages={071101},
  year={2011}, publisher={AIP Publishing},
  doi={10.1063/1.3610677}
}

@article{bennett1992quantum,
  title={Quantum cryptography without Bell's theorem},
  author={Bennett, Charles H. and Brassard, Gilles and Mermin, N. David},
  journal={Physical Review Letters}, volume={68}, number={5}, pages={557--559},
  year={1992}, publisher={American Physical Society},
  doi={10.1103/PhysRevLett.68.557}
}

@article{gerhardt2011full,
  title={Full-field implementation of a perfect eavesdropper on a quantum cryptography system},
  author={Gerhardt, Ilja and Liu, Qin and Lamas-Linares, Ant{\'\i}a and Skaar, Johannes and Kurtsiefer, Christian and Makarov, Vadim},
  journal={Nature Communications}, volume={2}, number={1}, pages={349},
  year={2011}, publisher={Nature Publishing Group},
  doi={10.1038/ncomms1348}
}

@article{tomamichel2012tight,
  title={Tight finite-key analysis for quantum cryptography},
  author={Tomamichel, Marco and Lim, Charles Ci Wen and Gisin, Nicolas and Renner, Renato},
  journal={Nature Communications}, volume={3}, number={1}, pages={634},
  year={2012}, publisher={Nature Publishing Group},
  doi={10.1038/ncomms1631}, note={arXiv:1103.4130}
}

@article{xu2020secure,
  title={Secure quantum key distribution with realistic devices},
  author={Xu, Feihu and Ma, Xiongfeng and Zhang, Qiang and Lo, Hoi-Kwong and Pan, Jian-Wei},
  journal={Reviews of Modern Physics}, volume={92}, number={2}, pages={025002},
  year={2020}, publisher={American Physical Society},
  doi={10.1103/RevModPhys.92.025002}, note={arXiv:1903.09051}
}

@article{stipcevic2013sap500,
  title={Characterization of a Commercially Available Large Area, High Detection Efficiency Single-Photon Avalanche Diode},
  author={Stip{\v{c}}evi{\'c}, Mario and Wang, Daqing and Ursin, Rupert},
  journal={Journal of Lightwave Technology}, volume={31}, number={23}, pages={3591--3596},
  year={2013}, publisher={IEEE},
  doi={10.1109/JLT.2013.2286422},
  note={arXiv:1311.2070}
}

@article{xu2024freerunning,
  title={Compact Free-Running {InGaAs/InP} Single-Photon Detector with 40\% Detection Efficiency and 2.3 kcps Dark Count Rate},
  author={Xu, Qi and Yu, Chao and Chen, Wei and Zhao, Jianglin and Cui, Dajian and Zhang, Jun and Pan, Jian-Wei},
  journal={IEEE Journal of Selected Topics in Quantum Electronics}, volume={30}, number={1}, pages={6400107},
  year={2024}, publisher={IEEE},
  doi={10.1109/JSTQE.2023.3328870},
  note={arXiv:2310.17113}
}

@article{chang2021detecting,
  title={Detecting telecom single photons with $99.5^{+0.5}_{-2.07}$\% system detection efficiency and high time resolution},
  author={Chang, J. and Los, J. W. N. and Tenorio-Pearl, J. O. and Noordzij, N. and Gourgues, R. and Guardiani, A. and Zichi, J. R. and Pereira, S. F. and Urbach, H. P. and Zwiller, V. and Dorenbos, S. N. and Esmaeil Zadeh, I.},
  journal={APL Photonics}, volume={6}, number={3}, pages={036114},
  year={2021}, publisher={AIP Publishing},
  doi={10.1063/5.0039772}
}

@misc{idq_id281,
  title={{ID281} Superconducting Nanowire Series},
  author={{ID Quantique}}, year={2025},
  note={Product web page, accessed 13 October 2026. System detection efficiency of at least 80--90\%, which can exceed 95\%; dark counts $<$1--100~cps at 1300--1600~nm; timing jitter 20--40~ps; recovery time $<$30~ns ($>$30~Mcps); 1 to 16 channels per cryostat},
  howpublished={\url{https://www.idquantique.com/quantum-detection-systems/products/id281-snspd-system/}}
}

@article{peng2025nondegenerate,
	title={Compact non-degenerate entangled-photon source and near-infrared-to-telecom quantum teleportation},
	author={Peng, Xu-Jie and Kong, Ling-Xuan and Lu, He},
	journal={Advanced Quantum Technologies},
	volume={8},
	number={11},
	pages={e00574},
	year={2025},
	doi={10.1002/qute.202500574},
	note={arXiv:2510.03620},
	publisher={Wiley}
}

@article{lydersen2011controlling,
  author  = {Lydersen, Lars and Akhlaghi, Mohsen K. and Majedi, A. Hamed and Skaar, Johannes and Makarov, Vadim},
  title   = {Controlling a superconducting nanowire single-photon detector using tailored bright illumination},
  journal = {New Journal of Physics}, volume = {13}, number = {11}, pages = {113042},
  year    = {2011}, doi = {10.1088/1367-2630/13/11/113042}
}

@article{hentschel2009three,
  author  = {Hentschel, Michael and H{\"u}bel, Hannes and Poppe, Andreas and Zeilinger, Anton},
  title   = {Three-color {Sagnac} source of polarization-entangled photon pairs},
  journal = {Optics Express}, volume = {17}, number = {25}, pages = {23153},
  year    = {2009}, doi = {10.1364/OE.17.023153}
}

@article{pelton2004bright,
  author  = {Pelton, Matthew and Marsden, Philip and Ljunggren, Daniel and Tengner, Maria and Karlsson, Anders and Fragemann, Anna and Canalias, Carlota and Laurell, Fredrik},
  title   = {Bright, single-spatial-mode source of frequency non-degenerate, polarization-entangled photon pairs using periodically poled {KTP}},
  journal = {Optics Express}, volume = {12}, number = {15}, pages = {3573},
  year    = {2004}, doi = {10.1364/OPEX.12.003573}
}

@article{natarajan2012superconducting,
  author  = {Natarajan, Chandra M. and Tanner, Michael G. and Hadfield, Robert H.},
  title   = {Superconducting nanowire single-photon detectors: physics and applications},
  journal = {Superconductor Science and Technology}, volume = {25}, number = {6}, pages = {063001},
  year    = {2012}, doi = {10.1088/0953-2048/25/6/063001}
}

@article{li2011attacking,
  author  = {Li, Hong-Wei and Wang, Shuang and Huang, Jing-Zheng and Chen, Wei and Yin, Zhen-Qiang and Li, Fang-Yi and Zhou, Zheng and Liu, Dong and Zhang, Yang and Guo, Guang-Can and Bao, Wan-Su and Han, Zheng-Fu},
  title   = {Attacking a practical quantum-key-distribution system with wavelength-dependent beam-splitter and multiwavelength sources},
  journal = {Physical Review A}, volume = {84}, number = {6}, pages = {062308},
  year    = {2011}, doi = {10.1103/PhysRevA.84.062308}
}

@article{neumann2022experimental,
  author  = {Neumann, Sebastian Philipp and Selimovic, Mirela and Bohmann, Martin and Ursin, Rupert},
  title   = {Experimental entanglement generation for quantum key distribution beyond 1 {Gbit/s}},
  journal = {Quantum}, volume = {6}, pages = {822},
  year    = {2022}, doi = {10.22331/q-2022-09-29-822}, note = {arXiv:2107.07756}
}

@misc{qoj_elvis,
  author  = {{Quantum Optics Jena GmbH}},
  title   = {{ELVIS Enterprise} Quantum Key Distribution System, Product Brief},
  note    = {Web version, received 13 October 2026. Source at 810 and 1550~nm, pair generation rate $>$25~MHz, Si and InGaAs SPAD receivers, typical key rate $>$500~bps at 10~dB loss},
  howpublished = {\url{https://www.qo-jena.com}}
}

@misc{qubitrium_qdcv,
  author  = {{Qubitrium}},
  title   = {{QDC-V} Series Single Photon Detector, Datasheet},
  note    = {Received 13 October 2026. Module QDC-V2-01: dark count 100~cps typical, timing jitter 0.3~ns at 300~kcps (up to 1~ns), dead time 60~ns, afterpulsing $<$0.1\%},
  howpublished = {\url{https://qubitrium.tech}}
}

@article{weier2011quantum,
  author  = {Weier, Henning and Krauss, Harald and Rau, Markus and F{\"u}rst, Martin and Nauerth, Sebastian and Weinfurter, Harald},
  title   = {Quantum eavesdropping without interception: an attack exploiting the dead time of single-photon detectors},
  journal = {New Journal of Physics}, volume = {13}, number = {7}, pages = {073024},
  year    = {2011}, doi = {10.1088/1367-2630/13/7/073024}
}

@article{lemos2014quantum,
  author  = {Lemos, Gabriela Barreto and Borish, Victoria and Cole, Garrett D. and Ramelow, Sven and Lapkiewicz, Radek and Zeilinger, Anton},
  title   = {Quantum imaging with undetected photons},
  journal = {Nature}, volume = {512}, pages = {409--412},
  year    = {2014}, doi = {10.1038/nature13586}, note = {arXiv:1401.4318}
}

@article{kalashnikov2016infrared,
  author  = {Kalashnikov, Dmitry A. and Paterova, Anna V. and Kulik, Sergei P. and Krivitsky, Leonid A.},
  title   = {Infrared spectroscopy with visible light},
  journal = {Nature Photonics}, volume = {10}, pages = {98--101},
  year    = {2016}, doi = {10.1038/nphoton.2015.252}, note = {arXiv:1506.07223}
}

@article{lagorivera2021telecom,
  author  = {Lago-Rivera, Dario and Grandi, Samuele and Rakonjac, Jelena V. and Seri, Alessandro and de Riedmatten, Hugues},
  title   = {Telecom-heralded entanglement between multimode solid-state quantum memories},
  journal = {Nature}, volume = {594}, pages = {37--40},
  year    = {2021}, doi = {10.1038/s41586-021-03481-8}
}

@article{lee2019symmetrical,
  author  = {Lee, Jianwei and Shen, Lijiong and Cer{\`e}, Alessandro and Troupe, James and Lamas-Linares, Antia and Kurtsiefer, Christian},
  title   = {Symmetrical clock synchronization with time-correlated photon pairs},
  journal = {Applied Physics Letters}, volume = {114}, number = {10}, pages = {101102},
  year    = {2019}, doi = {10.1063/1.5086493}, note = {arXiv:1812.08450}
}

@article{spiess2023clock,
  author  = {Spiess, Christopher and T{\"o}pfer, Sebastian and Sharma, Sakshi and Kr{\v{z}}i{\v{c}}, Andrej and Cabrejo-Ponce, Meritxell and Chandrashekara, Uday and D{\"o}ll, Nico Lennart and Riel{\"a}nder, Daniel and Steinlechner, Fabian},
  title   = {Clock synchronization with correlated photons},
  journal = {Physical Review Applied}, volume = {19}, number = {5}, pages = {054082},
  year    = {2023}, doi = {10.1103/PhysRevApplied.19.054082}, note = {arXiv:2108.13466}
}

@article{polyakov2007high,
  author  = {Polyakov, Sergey V. and Migdall, Alan L.},
  title   = {High accuracy verification of a correlated-photon-based method for determining photon-counting detection efficiency},
  journal = {Optics Express}, volume = {15}, number = {4}, pages = {1390},
  year    = {2007}, doi = {10.1364/OE.15.001390}
}

@misc{krelina2026quantum,
  author  = {Krelina, Michal and Tefek, Utku and Seskir, Zeki C. and Durak, Kadir},
  title   = {Quantum Protocols for Time Synchronisation and Distribution: A Critical Assessment},
  year    = {2026}, note = {arXiv:2604.10243},
  doi     = {10.48550/arXiv.2604.10243}
}

\end{document}